\documentclass[aps,rsi,floatfix,twocolumn,superscriptaddress,footinbib,reprint]{revtex4}

\usepackage[OT1,T1]{fontenc}
\usepackage[latin1]{inputenc}
\usepackage[english]{babel}
\usepackage[toc,page]{appendix}
\usepackage{textcomp}
\usepackage{moreverb}
\usepackage{setspace}
\usepackage{epsf}
\usepackage{graphicx}
\usepackage{graphics}
\usepackage{psfrag}
\usepackage{caption}
\usepackage{placeins}
\usepackage{amsmath}
\usepackage{amsfonts}
\usepackage{amssymb}
\usepackage{amstext}
\usepackage{mathrsfs}
\usepackage{anysize}
\usepackage{bm}
\usepackage[squaren,Gray]{SIunits}
\usepackage{hyperref}

\usepackage[RGB]{xcolor} 
\definecolor{darkblue}{rgb}{0,0,.4}

\definecolor{darkgreen}{rgb}{0,.4,0}
\newcommand{\Cred}[1]{{\color{red} #1}}
\newcommand{\Cgreen}[1]{{\color{darkgreen} #1}}
\newcommand{\Cblue}[1]{{\color{blue} #1}}
\newcommand{\kar}{von K\'{a}rm\'{a}n}
\newcommand{\fprop}{f_\text{prop}}
\newcommand{\eg}{\emph{e.g.}}
\newcommand{\ie}{\emph{i.e.}}
\renewcommand{\vec}{\bm}
\providecommand{\abs}[1]{\lvert#1\rvert}

\hypersetup{pdftex=true, colorlinks=true, breaklinks=true, linkcolor=darkblue, menucolor=darkblue,  urlcolor=darkblue, citecolor=darkgreen}

\begin{document}

\title{Measuring Lagrangian accelerations using an instrumented particle}

\author{R.~Zimmermann}
\affiliation{Laboratoire de Physique, ENS de Lyon, UMR CNRS 5672, Universit\'e de Lyon, France}
\email{robert.zimmermann@ens-lyon.org}
\author{L.~Fiabane}
\affiliation{Laboratoire de Physique, ENS de Lyon, UMR CNRS 5672, Universit\'e de Lyon, France}
\author{Y.~Gasteuil}
\affiliation{smartINST S.A.S., 46 all\'ee d'Italie, 69007 Lyon, France}
\author{R.~Volk}
\affiliation{Laboratoire de Physique, ENS de Lyon, UMR CNRS 5672, Universit\'e de Lyon, France}
\author{J.-F.~Pinton}
\affiliation{Laboratoire de Physique, ENS de Lyon, UMR CNRS 5672, Universit\'e de Lyon, France}

\begin{abstract}
Accessing and characterizing a flow impose a number of constraints on the employed measurement techniques; in particular  optical methods require transparent  fluids and windows in the vessel. Whereas one can adapt apparatus, fluid and methods in the lab to these constraints, this is hardly possible for industrial mixers.
We present in this article a novel measurement technique which is suitable for opaque or granular flows: an instrumented particle, which continuously transmits the force/acceleration acting on it as it is advected in a flow. Its density is adjustable for a wide range of fluids and because of its small size and its wireless data transmission, the system can be used both in industrial and scientific mixers allowing a better understanding of the flow within. 
We demonstrate the capabilities and precision of the particle by comparing its transmitted acceleration to alternative measurements, in particular in the case of a turbulent von K\'arm\'an flow. Our technique shows to be an efficient and fast tool to characterize flows.
\end{abstract}

\date{\today}
\maketitle

\section{Introduction}

Experimental fluid dynamics research in the lab consists of an interplay of  suitable flow generation devices, working fluids, measurement techniques and analysis, with goals ranging from fundamental research in statistical / non-linear physics to the optimization of mixers in industrial R\&D departments.
 In this endeavor, very significant progress has been achieved during the last decade with the advent of space and time resolved optical techniques based on high speed imaging~\cite{book:expFluids}. 
However, direct imaging is not always possible especially in industry: opaque vessels, non-transparent fluids, environmental constraints among other things may be limiting factors. 
Even if the fluid is transparent, the injection of tracer particles might be still not allowed or unsuitable  due to bio-medical or food regulations, or due to the chemical properties of the fluid. 
While techniques using other kinds of probing waves (\eg{} acoustics~\cite{Mordant:2001ts}) have been developed, a direct resolution of the Eulerian flow pattern is not always possible. 
In this context, Lagrangian techniques provide an interesting alternative  particularly for problems related to mixing~\cite{annRev:LagProps,Shraiman:2000vu}. 

Lagrangian tracers with a temperature sensitive dependance have been used in the study of Rayleigh-B\'enard convection~\cite{Xia:2012}, a problem for which our group has developed  \emph{instrumented particles}~\cite{Gasteuil:2007tw,Gasteuil:2009la,Shew:2007cy,Pinton:fk}. 
The approach was to instrument a neutrally buoyant particle in such a way that it  measures the temperature fluctuations during its motion as it is entrained by the flow, while transmitting the data via radio frequency to a lab operator in real time. 
Meaningful information regarding the statistics of thermion plumes have been obtained, with excellent agreement with other techniques~\cite{Xia:2012} and direct numerical simulations~\cite{Schumacher:2009jj}. 
In the work reported here, we built upon this approach to instrument the particle such that one gets flow parameters directly from the measurements (in \cite{Gasteuil:2007tw}, one had to simultaneously film the particle motion). 
We equip the particle with a 3-axis accelerometer, whose measurements are sampled at a rate equal to $316$~Hz and transmitted to the lab operator. 
This particle is intended for turbulent flows. Thanks to its radio transmission it is suitable for opaque fluids or apparatuses without access for optical measurement techniques. Its continuous operation is also advantageous over Particle Tracking Techniques which have to operate in chunks as the memory of the tracking cameras  is necessarily limited. Moreover and  in contrast to tracer particles this instrumented particle can be easily re-extracted from the apparatus after the experiment.
However, as the particle is advected in a flow it rotates and consequently continuously changes its  orientation with respect to the laboratory frame. 
Thereby the signals of the 3D accelerometer are altered in a non-trivial way, and detailed characterization and methods to extract meaningful information from the acceleration signals are needed.
We present here the preliminary results of this characterization.

This article is organized as follows:
first, we present the instrumented particle and additional techniques needed for its characterization (section~\ref{sec:smartpart}).
In section~\ref{sec:signals}, we present an analysis of the results obtained in two different configurations: 
First, a simple pendulum with the particle attached at the end of a stiff arm, then the particle advected in a fully turbulent flow.
In order to verify that the transmitted acceleration is well related to its motion,  we compare the results to simultaneous alternative measurements.
Finally, we discuss  limitations and perspectives of this new measurement technique (section~\ref{sec:discussion}). 

\begin{figure*}[tb] 
\centering
      \includegraphics[width=.8\textwidth]{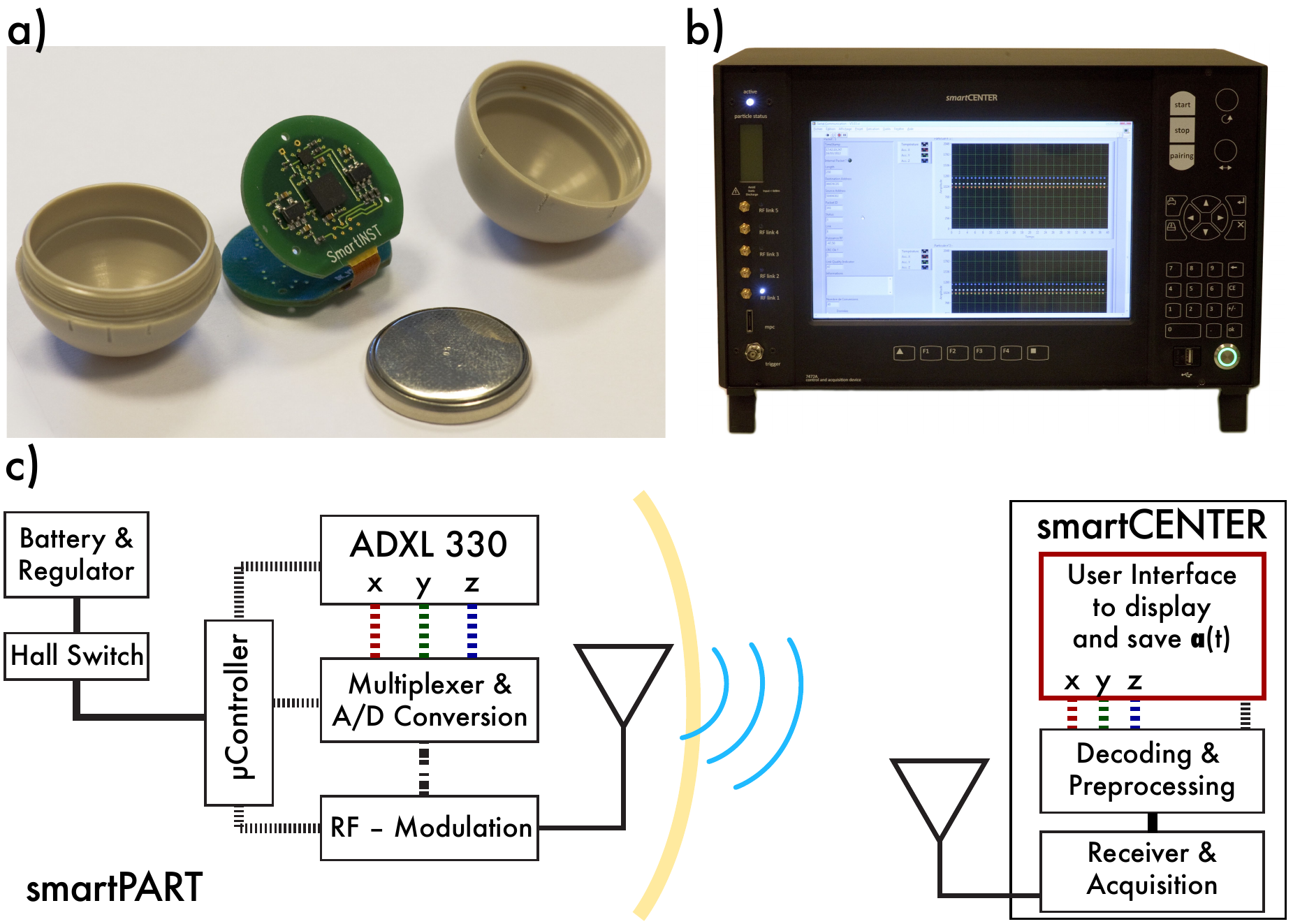} 
   \caption{\small  a) and b) : the instrumented particle (\emph{smartPART}) and its data control \& acquisition unit (\emph{smartCENTER}). The coin cell is  $20\milli\meter$ in diameter. \quad c)    The  diagram sketches the measurement of the  acceleration, its transmission to and the processing by the smartCenter.
   }
   \label{fig:smartPartCircuit}
\end{figure*}

\section{``Smart particles''}
\label{sec:smartpart}

The apparatus described in the following is designed and built by smartINST S.A.S., a young startup situated on the ENS de Lyon campus.  
The device consists of a spherical particle (the so-called smartPART) which embarks an autonomous circuit with 3D-acceleration sensor, a coin cell and a wireless transmission system; and a data acquisition center (the so-called smartCENTER), which acquires, decodes, processes and stores the signal of the smartPART (see Fig.~\ref{fig:smartPartCircuit}). 
The ensemble -- smartPART and smartCENTER -- measures, displays and stores the three dimensional acceleration vectors acting on the particle as it is advected in the flow.
The accelerations are observed  in a moving and rotating coordinate system and consist of four contributions: gravity,  translation, noise and possibly a  weak contribution of the rotation around the center of the particle itself.

\subsection{Design \& Technical Details}

\paragraph{Sensor:} 
The central component of the particle is the ADXL 330 (Analog Device) -- a three axis accelerometer. 
This component belongs to the category of micro-electro-mechanical systems (MEMS). Each of the three axes  returns  a voltage  proportional to the force acting on a small, movably mounted mass-load suspended by micro-fabricated springs. 
The three axes of the ADXL 330 are decoupled and form an orthogonal coordinate system attached to the chip package. 
From this construction arises a permanent measurement of the gravitational force/acceleration $\vec g\equiv 9.8 \, \meter/\second^2 \cdot\vec {\hat e}_g= g\cdot  \vec {\hat e}_g$. 
Each axis has a guaranteed minimum full-scale range of $\pm 3 \,g$; however, we observe a typical range of $\pm 3.6\, g= 35 \, \meter/\second^2$ per axis. 
The sensor has to be calibrated to compute the physical accelerations from the voltages of the accelerometer. \\

\paragraph{smartPART:}
The signals from the ADXL 330 are first-order low-pass filtered at $f_c=160\,\hertz$ and then digitized at  $12$ bits and $316\,\hertz$  sampling rate. A multiplexer prior the signal digitization  induces a small time shift between the components of $0.64\,\milli\second$. 
The output is then reshaped  into small packets  and send via radio frequency.
The ensemble  is powered by a coin cell. A voltage regulator  ensures a stable supply voltage and thus a constant quality of the measurement. 
A Hall switch allows one to power-down most components; thereby, the battery is only used during experiments.
Depending on the power needed to transmit the acceleration signals, a particle operates continuously for $6$ to $36$ hours. 
The ADXL 330 is soldered to the printed circuit board such that it is situated close to the geometrical center  of the particle. The particle itself is spherical with a diameter of $\unit{25}\,{\milli\meter}$. 
The capsule walls are made of Polyether-ether-ketone (PEEK) which is known for its excellent mechanical and chemical robustness. 
It is leak-proof and its density can be matched to fluids by adding extra weight (namely Tungsten paste) to the particle's interior; 
within the density range of  $\unit{0.8 - 1.4}\,{ \gram/\centi\meter^3}$ a relative density match of  better than $10^{-4}$ is achievable. The particle is thus suited for most experiments in water and water-based solutions.
It  should be noted that the mass distribution inside the particle is neither homogeneous nor isotropic: in particular its center of mass does not coincide with the geometrical center, making it out-of-balance. In practice this results into a  pendulum-like motion of the particle in the flow. Nevertheless, the imbalance can be adjusted to some extent  by adding patches of Tungsten paste to its interior, and the particles we use are carefully prepared such that they are neutrally buoyant,  avoid any pendulum-like behavior and rotate easily in the flow.  \\

\paragraph{smartCENTER:}
The signals from the smartPART are received by an antenna connected to the smartCENTER, which contains    radio reception, processing and display units. It  demodulates and decodes in real-time the received raw signal into a time-series of  raw voltages of the ADXL~330. The physical acceleration sensed by the smartPART $\vec a_\text{SP}$ can then be computed:
\begin{equation}
\vec a_\text{SP} = \left(\begin{array}{c}a_1 \\a_2 \\a_3 \end{array}\right)=\left(\begin{array}{c}(A_1-O_1) / S_1 \\(A_2-O_2)/ S_2 \\(A_3-O_3) / S_3 \end{array}\right),
\label{eq:rawacc}
\end{equation}
where  $ A_i $, $O_i$  and $S_i$ are   the measured raw signal, the offset and the sensitivity of each axis, respectively.
Offset and sensitivity have to be calibrated beforehand; the procedure is described in the following section.
The resulting time-series are saved for further processing.

\subsection{Calibration and robustness}

The offset and sensitivity of the ADXL 330 have to be calibrated to convert the measured voltages into a physical acceleration. 
The axes of the accelerometer form an orthogonal coordinate system according to Eq.~\eqref{eq:rawacc}. 
At rest one observes only gravity  projected onto the sensor at an arbitrary orientation. The observed raw values  define consequently a translated ellipsoid (for simplicity we set $\abs{\vec g}\equiv 1$):
 \begin{equation}
\vec a_\text{SP} \cdot \vec a_\text{SP} =\sum_i\frac{(A_i-O_i)^2}{S_i^2} = \vec g^2=1\;.
\label{eq:ellipCapt}
\end{equation}
Eq.~(\ref{eq:ellipCapt}) can be arranged to:
 \begin{equation}
 1=\sum_i{\left(\xi_i\,A_i^2-2\,\xi_{i+3}\,A_i\right)} \;,
\end{equation} 
with $\xi_i$ six parameters  containing  offset and sensitivity. 
A sufficient number of measurements with different orientations define a set of equations which is solved  using a linear least squares technique.  Offset and sensitivity are then 
\begin{equation}
O_i=\frac{\xi_{3+i}}{\xi_i} \quad \text{and}\quad  S_i= \sqrt{\frac{1+  \sum_i\big({\xi_{3+i}^2/\xi_{i}}\big)}{\xi_i}} ~~\text{.}
\end{equation}

We find that the particle at rest has an average noise of $\sigma_x =\sigma_y= 0.006\,g$ and $ \sigma_z=0.008\,g$, giving $\abs{\vec\sigma}=\sqrt{\sum_i\sigma_i^2}=0.012\,g$, where $g$ is the magnitude of gravity. 
An analysis using the residuals showed a slightly higher resolution of $\sigma_x =\sigma_y= 0.005\,g$ and $\sigma_z=0.003\,g$, and $\abs{\vec\sigma}=0.008\,g$.
These values are thus the absolute errors of our measurement. 

The ADXL~300 has among other things been chosen for its weak temperature dependance: its offset typically varies by $10^{-3} \,g/\degreecelsius$, and its sensitivity by  $0.015\,\%/\degreecelsius$. Digitizing and transmission unit were verified to be temperature independent. Consequently, the total temperature dependence of the smartPART is  given by its accelerometer. For high precision measurements, it is advised to calibrate the particle at experiment temperature shortly before the actual experiment.

We noticed a small drift of the order of $0.005\,g/\hour$ for  the $z-$axis. No drift was observed for  the $x-$ and $y-$axes. Since a voltage regulator ensures a stable supply voltage of the circuit, this drift stems most likely  from the internal construction of the accelerometer. 
Owing to the continuous data transmission of the instrumented particle, one flow configuration can be characterized in approximately 30 minutes. Hence, the little drift of the $z-$axis can be neglected.

Considering the mechanical robustness, the smartPART survived several days in a \kar{} mixer and neither contacts with the wall nor with the sharp edged blades of the fast rotating propellers damaged its function or shell.

\section{Acceleration signals}
\label{sec:signals}

As mentioned previously, the smartPART  transmits in real-time the accelerations acting on the particle as it is advected in the flow. 
The noise-to-signal ratio being small, we neglect the noise from here on.
The contributions consist therefore of: gravity,  translation, and rotation of the particle. 
We now test the accuracy of the particle signals in two different experimental configurations by comparison with alternative measurements.

\subsection{2D pendulum}
A pendulum is a simple and well-known case, ideal to measure the resolution of the particle.
A stiff pendulum with $60\,\centi\metre$ long stiff arm is equipped with a position sensor returning the deflection angle $\varphi$ of the arm. 
The particle is fixed at known length, $l$, with a known arbitrary orientation to the arm. The fact that a rotation of the particle around its center is restricted, implies that the contribution from the rotation of the particle around is center is known.
Measuring $\vec a_\text{SP}$ at rest ($\varphi=0\degree$) and at several arbitrary positions one can determine the axis of rotation of the arm.  
Once this vector is known  the measured acceleration signal is rotated/re-expressed such that $a_y$ points with the arm, $a_x$ with the movement, and $a_z$ with the axis of rotation. 
Note that by definition the latter does not change when the pendulum moves.
The  signal seen by the particle is a two-dimensional problem and fully described as a function of  the deflection angle $\varphi$:\\
\begin{equation}\begin{split}
a_x(\varphi)&=g \, \sin \varphi  +l \, \ddot \varphi\, ,\\
 a_y(\varphi)&=g \,  \cos \varphi + l \, {\dot\varphi}^2 \,.
\end{split}\label{eq:pendulum}
\end{equation}
In the limit of the small angle approximation, this simplifies to the well-known oscillations of frequency $\omega$:
\begin{equation}\begin{split}
a_x(\varphi)&\approx -l\omega^2 \, \sin{\omega t} \, ,\\
 a_y(\varphi)&\approx g+ \frac{l\,\omega^2}{2}(1+ \cos{2\omega t}) \,.
\end{split}\label{eq:pendulum2}\end{equation}
The simultaneous measurement of  the angle, $\varphi(t)$, and the particle's signal,  $\vec a_\text{SP}(t)$, enables us to compare the two signals without any other approximation or fit than Eq.~(\ref{eq:pendulum}). 
\begin{figure}[tb] 
   \centering
   \includegraphics[width=\columnwidth]{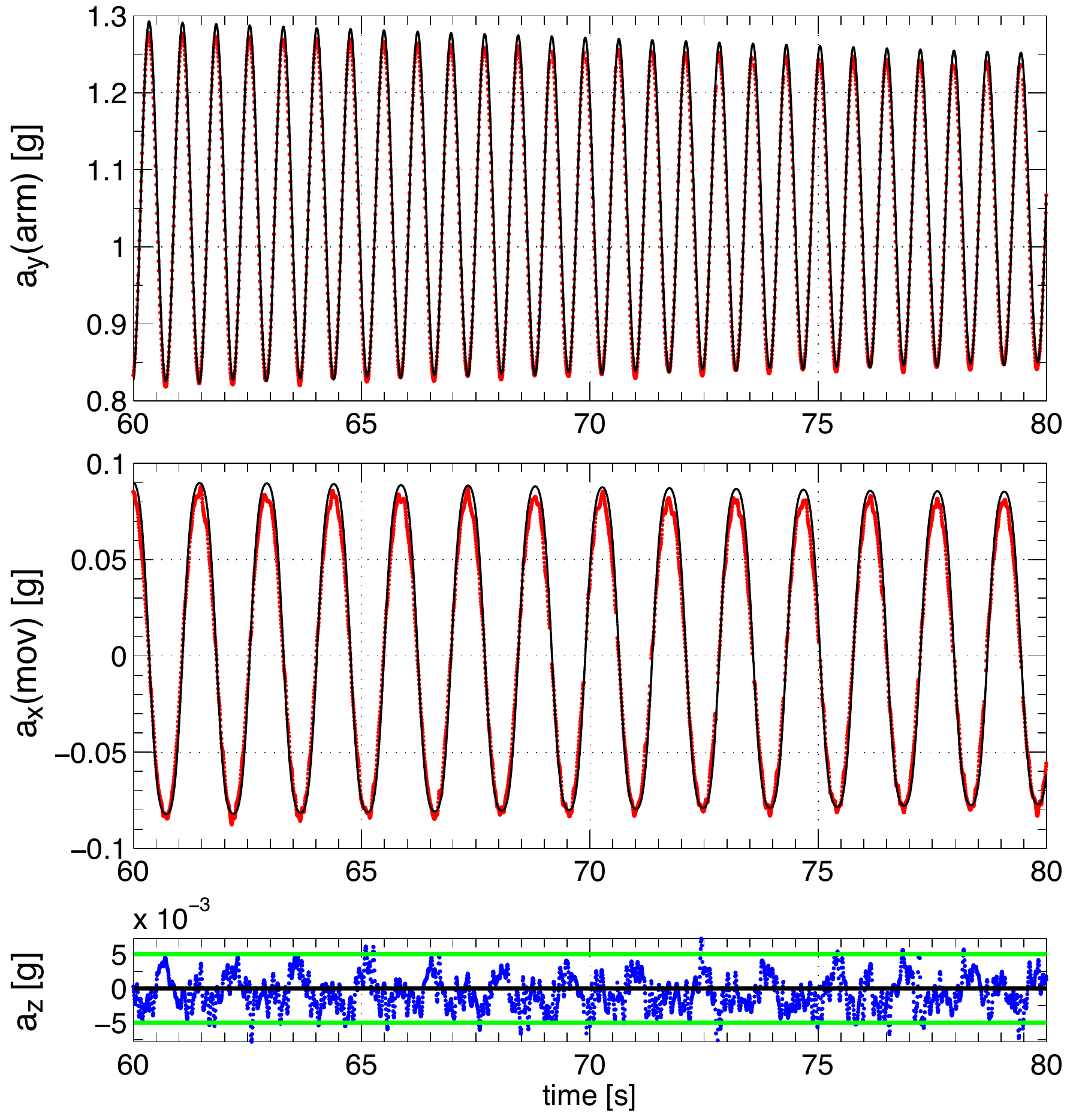} 
   \caption{\small Comparison of the particle's signal $ \vec a_\text{SP}(t)$ (\Cred{$\bullet$}) to the theoretical curves based on the position sensor (---). $a_y$  points with the arm and  $a_x$ measures the force in direction of the movement. No force is exerted  along the $z-$axis (the green lines represent the uncertainty of the calibration). Note, that acceleration is measured in $g=9.8\,\metre\per\second\squared$ . }
   \label{fig:pendel}
\end{figure}

Fig.~\ref{fig:pendel} shows $\vec a_\text{SP}(t)$ for several periods of the pendulum, measured by the smartPART and by the position sensor. 
The agreement between the two signals is very good, and in particular better than the uncertainty of the calibration.
Hence, the Lagrangian acceleration of the smartPART corresponds well to its actual motion in this simple case.
We now move on to more complicated motions.

\begin{figure*}[p] 
\centering
      \includegraphics[width=.8\textwidth]{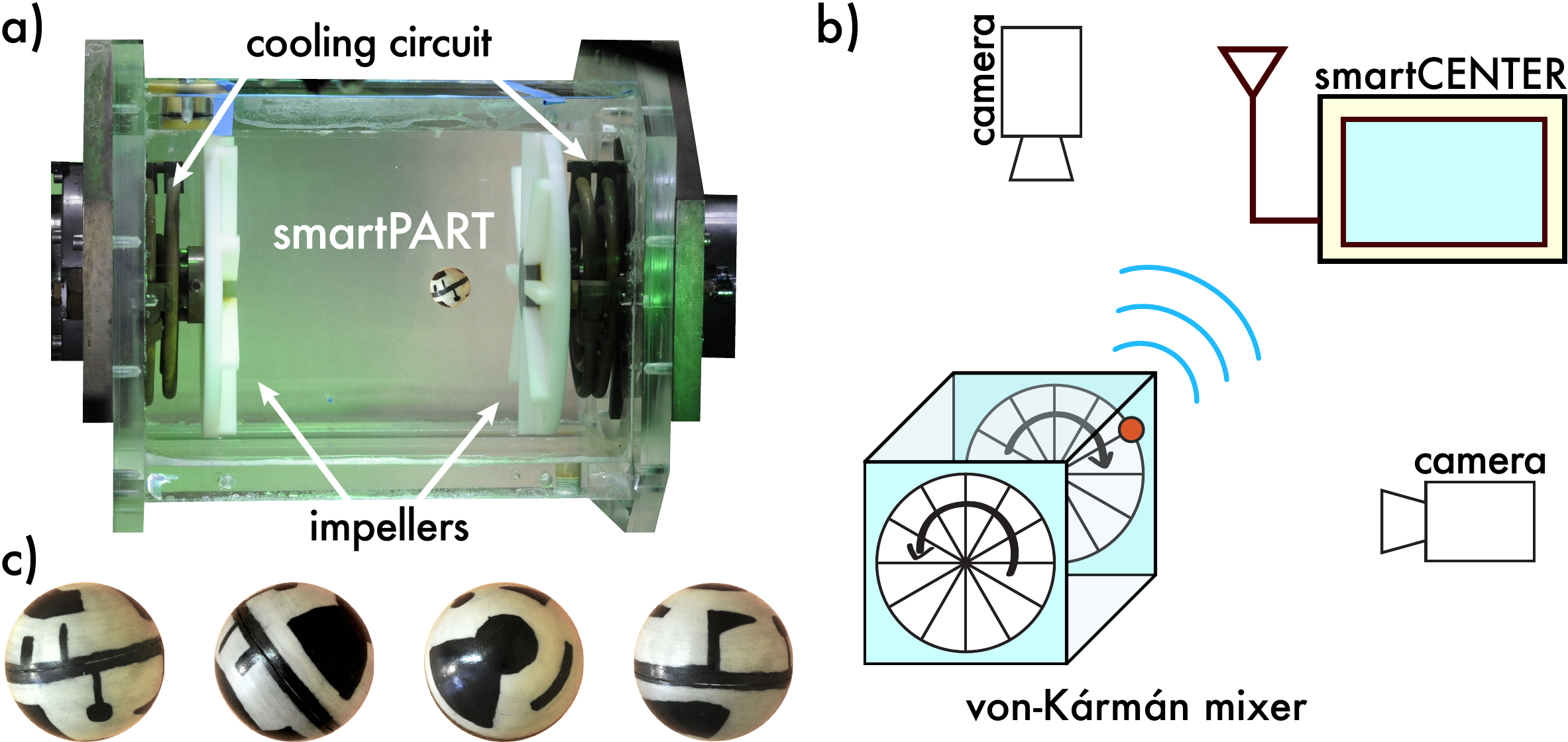} 
   \caption{\small Experimental setup: \quad a) picture of the apparatus; b) sketch of the arrangement;  c) a textured instrumented particle at different orientations.}
   \label{fig:vankarman}
\end{figure*}

\begin{figure*}[p] 
   \centering
   \includegraphics[width=\textwidth]{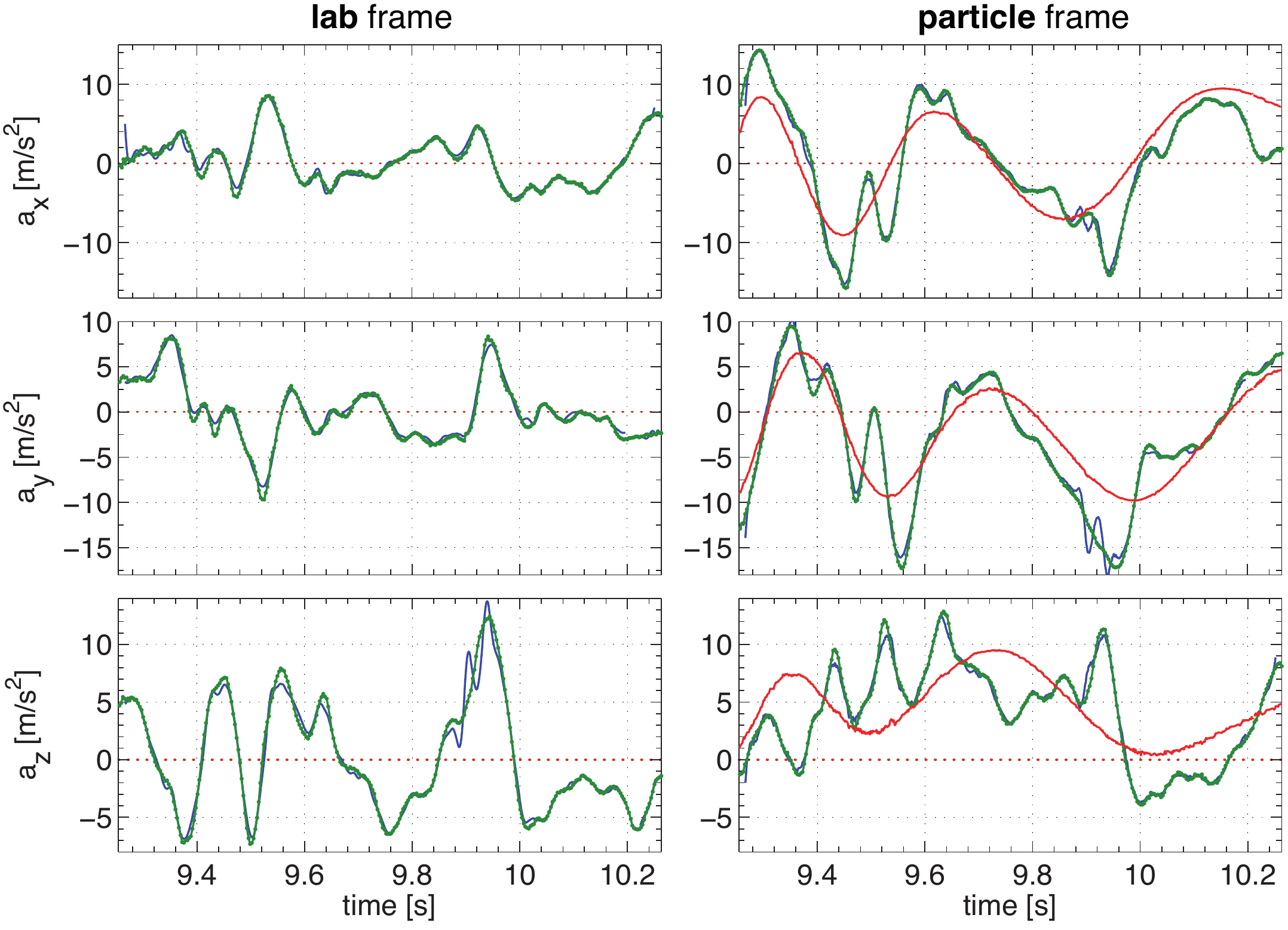} 
   \caption{\small A sample trajectory  of the instrumented particle seen by the camera (\Cblue{---}) or smartPART (\Cgreen{$\bullet$}), it is $\fprop=3\,\hertz$.  The absolute orientation enables us to re-express the camera measurement   of  the particle (lab  frame) in the moving  frame of the particle and vice-versa. In the former  gravity is subtracted and in the particle frame gravity is represented by the \Cred{red} line.  }
   \label{fig:combJimCam}
\end{figure*}

\subsection{Fully developed (3D) turbulence}
The instrumented particle is intended for the characterization of complex/turbulent flows. Such flows exhibit strong, intermittent variations in the acceleration. To verify the  suitability of the smartPART for these conditions, we now investigate its motion in a fully turbulent mixer while tracking it with an independent optical technique.

Namely, we use a \kar{} water flow: a swirling flow is created in a square tank by two opposing counter-rotating impellers of radius $R=9.5\,\centi\meter$ fitted with straight blades $1\,\centi\meter$ in height. 
The flow domain in between the impellers has characteristic length $H  = 20\,\centi\meter \cong 2R$ (see Fig.~\ref{fig:vankarman}) and the vessel is built with transparent  flat  side walls, allowing direct optical measurements over almost the whole flow domain. 
Blades on the impellers work similar to a centrifugal pump and add a poloidal circulation at each impeller.  
For counter-rotating impellers, this type of flow is known to exhibit fully developed turbulence~\cite{RAVELET:2008ja}.
Within a small region in the center  the mean flow is little and  the local characteristics approximate homogeneous turbulence. However, at  a large scale  it is known to have a large scale anisotropy~\cite{Ouellette:2006fk,Monchaux:2006fj}. At a propeller frequency of $3\,\hertz$ we estimate a   Reynolds number based on the Taylor micro-scale of $R_\lambda = 500\pm50$.

We optically track the translation and absolute orientation of the smartPART while simultaneously acquiring the  transmitted acceleration time-series. These optical measurements are then used as a reference to compare with the instrumented particle's signals. 
The six-dimensional tracking technique (or 6D tracking,  3 components for the translation and 3 components for the rotation of the particle around its center) is explained in detail in~\cite{Zimmermann:2011uu,Zimmermann:2011fk} and briefly sketched here (Fig.~\ref{fig:vankarman}).
In order to determine  the absolute orientation the particle is textured by hand using black-ink permanent marker (see Fig.~\ref{fig:vankarman}c).   Acceleration sensor and texture are then calibrated/retrieved independently; nevertheless,  the accelerometer is at a fixed but unknown orientation with respect to this texture, \ie{} sensor and texture are related by a constant rotation matrix. We determine this matrix by  acquiring the acceleration signals of the particle at arbitrary orientations while additionally determining its orientation and the location of gravity on the texture. The particle is then inserted into the apparatus, which is illuminated by high power LEDs. Its motion is tracked by two high-speed video cameras (Phantom V12, Vision Research) which record synchronously two views at approximately 90 degrees.  The observation volume is  $15\times 15\times 15 \; [\centi\metre\cubed]$ in size and resolved  at a resolution of $4.2\,{\text{pixel}}\per{\milli\metre}$. In our configuration, a camera can store on the order of $14\,000$ frames in on-board memory, thus limiting the duration of continuous tracks. Therefore, a computer  issues  the  recording of 8 bit gray-scale movies  at a sufficiently high frame rate while controlling  the smartCENTER such that acceleration signal and images are synchronized. After extracting the time-series of the particle's position and orientation,  one can then compare the accelerometer's signal to the motion of the particle.

It should be stressed that the two measurement techniques observe the motion of the instrumented particle in two completely different reference frames. 
On the one hand, the 6D tracking uses a fixed, non-rotating coordinate system, and is referred to as the lab frame.
On the other hand, as the particle is advected and turned in the flow, it and consequently the embarked accelerometer   constantly rotate their coordinate system with respect to the lab frame; the acceleration signal is thus measured in a  frame which is continuously rotating and not fixed. This frame is referred to as the particle frame.
The acceleration sensor measures the forces acting on it as it moves in the flow. Knowing the absolute orientation of the particle at each instant we can  express  the signal of the smartPART in the lab frame by rotating it such that it  corresponds to a non-rotating particle. Starting from the time-series of position and orientation, it is also possible to compute the linear, centrifugal and gravitational acceleration/force  acting on a point inside the particle and then project these into the rotating particle frame. The different components are then expressed in the frame of the sensor. 

Fig.~\ref{fig:combJimCam} shows a sample trajectory in both coordinate systems. The agreement between the two techniques is remarkable. Furthermore, one observes that the projection of gravity is continuously changing: the particle is rotating in a non-trivial way. Deviations between the two techniques stem from several experimental errors.
First, the position measurement: bubbles, reflections and other impurities  alter the measured position of the particle. The acceleration is the second derivate and thus highly sensitive to such events.
Second, the orientation measurement: the absolute orientation is needed to change between the reference frames. The uncertainty in the absolute orientation is typically $3\degree$; that results in a wrong projection of gravity of $\pm0.5\,\meter/\second^2$. It further biases the rotational forces, as they are derivatives of the orientation time-series.
Finally, the matrix relating sensor and texture: this matrix is constant and thus a systematic contribution. The uncertainty is less than $2\degree$ -- \ie{} the error in projecting  gravity is  $<0.3\, \meter/\second^2$.
The observed agreement in the lab frame, $\Delta \vec a=\vec a_\text{SP}-\vec a_\text{6D}$, between the two techniques is as follows:  all three components of $\Delta \vec a$ have the same PDF. Surprisingly, the (absolute) uncertainty almost doubles by increasing $\fprop$ from $2\,\hertz$ to $3\,\hertz$. Nevertheless, for $80\%$ of the data the agreement is better than $0.8\,\meter/\second^2$ and $1.6\,\meter/\second\squared$, respectively. For comparison,  the absolute value $\abs{\vec a_\text{6D}}$ has a mean of  $2.9\,\meter/\second\squared$ and  $6.6\,\meter/\second\squared$ and a standard deviation of $1.8\,\meter/\second\squared$ and  $4.1\,\meter/\second\squared$, respectively.
The signal of the particle is thus corresponding to the flow, however, its interpretation is not simple. In particular, and after comparing many different trajectories, it becomes clear that no easy transformation is available to get rid of the rotation of the particle.

\begin{figure}[tb] 
   \centering
   \includegraphics[width=0.48\textwidth]{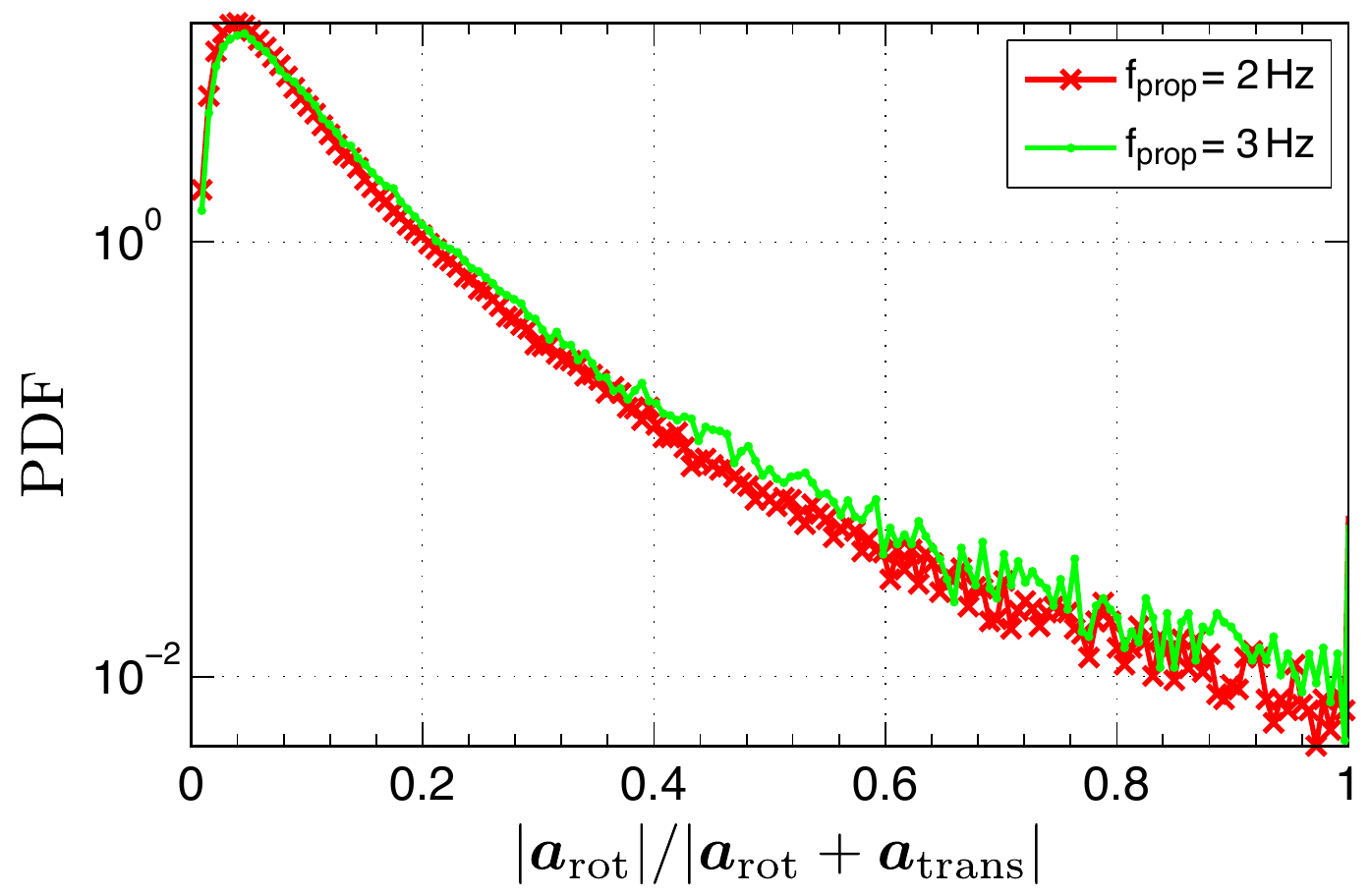} 
   \caption{\small Ratio of the rotational forces to the total force acting on the particle. The $80\%$ percentile is found at a ratio of $0.14$ and $0.16$, respectively.}
   \label{fig:forceRatio}
\end{figure}

By construction the center of the accelerometer is placed at $\vec r = 3\,\milli\meter\,\cdot \hat{\vec e}_z$.
A rotation of the particle around its geometric center  will thus add a centrifugal contribution to the measured acceleration.
This rises the question which term -- translation or rotation of the particle -- dominates the acceleration signal. To address this question we take advantage of the 6D-tracking, which enables us to compute the different forces acting on a point at $\vec r = 3\,\milli\meter\,\cdot \hat{\vec e}_z$ inside the sphere.  We can thus  compare the contribution of the translation and that of the rotation of the particle.
Fig.~\ref{fig:forceRatio} shows the ratio of the rotational (\ie{} centrifugal)  acceleration, $\vec a_\text{rot}=\vec\omega\times\vec\omega\times\vec r + \frac{\text{d}\vec\omega}{\text{d}t}\times\vec r$ to the total acceleration, $\vec a_\text{trans}+\vec a_\text{rot}$, (without gravity).  
Dimensional arguments tell that $a_\text{trans}\propto \fprop^2$ and $a_\text{rot}\propto \fprop^2$. 
Consistently, the PDF of the ratio $\abs{\vec a_\text{rot}}/\abs{\vec a_\text{trans}+\vec a_\text{rot}}$ differs only little for the two   propeller frequencies. 
Moreover, it is  peaked at $5\%$ and the $80\%$ percentile is at a ratio of $14\%$ and $16\%$, respectively. 
Hence, it is legitimate to neglect the rotational forces if no 6D tracking is available.
 
\section{Discussion}
\label{sec:discussion}

\begin{figure}[tb] 
   \centering
   \includegraphics[width=.48\textwidth]{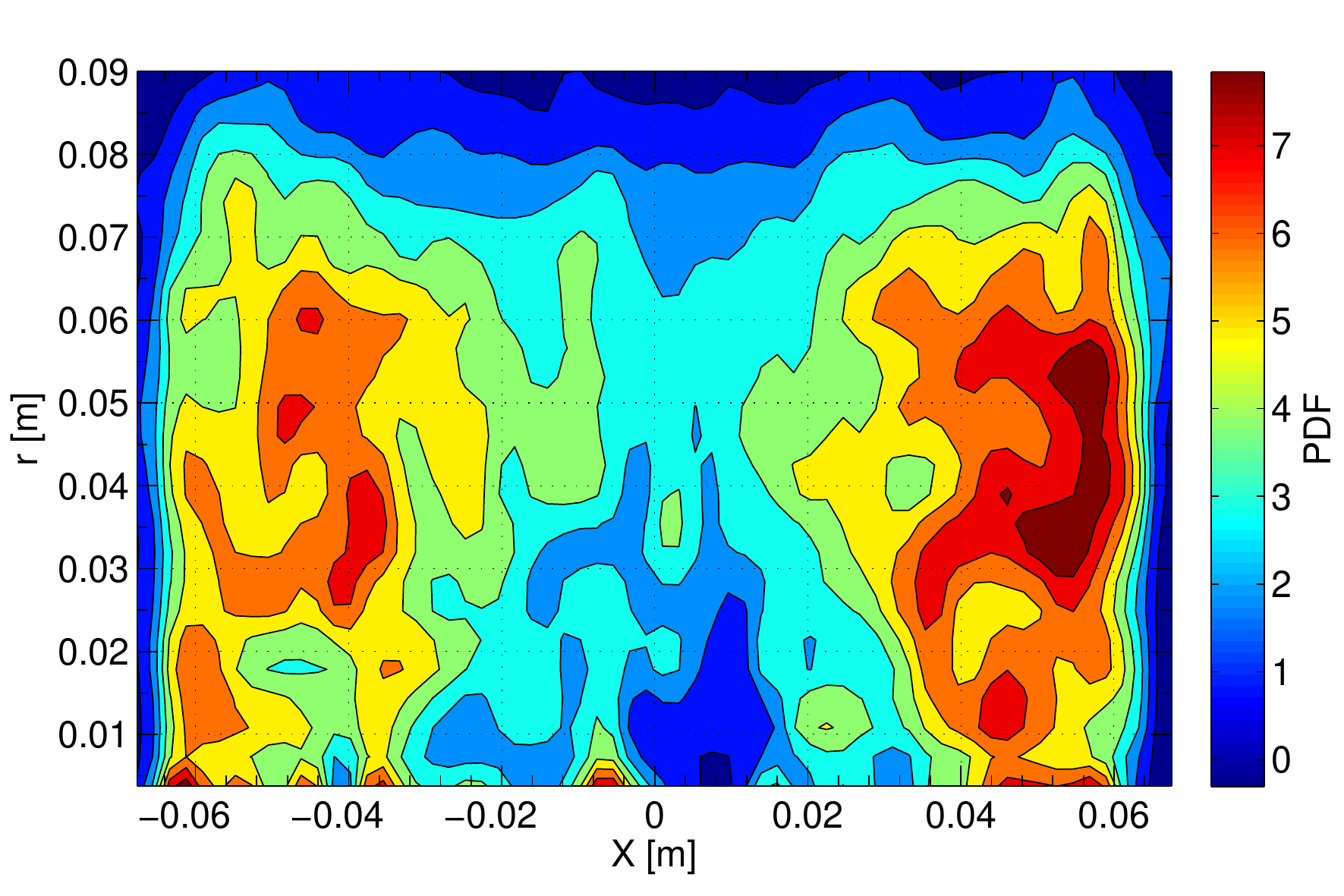} 
   \caption{\small Preferred position of the instrumented particle: independent of the propeller speed it is mostly situated in a torus shape around the propeller.}
   \label{fig:jimPosPDF}
\end{figure}

In the latter part of this article we studied the behavior of a large neutrally buoyant sphere in a turbulent flow.
Comparing with solid spheres of the same size in the same mixer, we find that the particle in general behaves almost identically~\cite{Zimmermann:2012fk}. 
In particular (and despite the fact that the instrumented particle is neutrally buoyant) we observe that it generally stays in a region close to the impellers. 
Fig.~\ref{fig:jimPosPDF} shows the PDF of position for the smartPART. 
Independent of the impeller speed it is mostly situated in a torus shape around the propeller, exhibiting a preferential  sampling of the flow for these large neutrally buoyant spheres. 

Moreover, since we investigate large particles with a size $D_\text{part}$ comparable to the integral length scale, $L_\text{int}$, moving through the whole mixer, the Kolmogorov assumptions to characterize turbulence are no longer valid. 
For these reasons, the smartPART can be insufficient to access all details of  a turbulent flow: some parts of the flow are little explored, and some scales of the turbulence might be filtered due to the size of the instrumented particle. 
However, one should bear in mind that these features of the flow are often accessed by means of optical methods whereas  the instrumented particle operates also in environments  and fluids which are unsuitable for optical measurement techniques.

Some other experimental constraints should be additionally stressed here.
As previously said, the mass distribution inside the particle is neither homogeneous nor isotropic. 
It is therefore possible that the particle is out-of-balance, \ie{} that the center of mass does not coincide with the geometrical center. 
Such a particle has a strong preferred orientation and wobbles similar to a kicked physical pendulum. 
The imbalance can be adjusted to some extent by adding weight to its interior, but the particle must be prepared very carefully and one must make sure that the particle used is well-balanced and rotates easily in the flow.

Also, the receiver/demodulation unit  of the smartCENTER works best within a range of  radio power, \ie{} particles which are emitting either too strong or too weak are undesirable and one has to adjust the radio emission of the smartPART.
A stronger radio emission power can  be required, \eg{} if the apparatus builds a Faraday cage (\ie{} an electrically-connected metal structure surrounds the flow), or if the signal has to pass a longer distance in more water or in a bigger apparatus. Solutions with a high conductivity are also likely to  damp the radio signal.
Naturally, a stronger radio emission shortens the life time of the battery.
Nevertheless, particles with stronger radio emission still last $6$ to $12$ hours, which is  sufficient in most cases.\\

To conclude, we presented the working principle of an instrumented particle giving a measure of the three components of the Lagrangian acceleration.
We were able to show that the Lagrangian acceleration of the smartPART corresponds well to its actual translation {and is not biased by a possible rotation of the particle around its center}. 
Work on extracting detailed information on the flow from the acceleration time-series is ongoing. 
This instrumented particles can shed some light into mixers which were not or hardly accessible up to now. Due to its continuous transmission  one flow configuration can be characterized within $\sim30\,\minute$.  
Apart from its appeal for chemical and pharmaceutical industry, it might  be an interesting tool to quantify  flows  in  labs. \\

\begin{acknowledgments}
This work was partially supported by ANR-07-BLAN-0155.  The authors want to acknowledge the technical help of Marius Tanase and Arnaud Rabilloud for the electronics, and of all the ENS machine shop.
 The authors also thank Michel Vo{\ss}kuhle, Micka\"el Bourgoin and Alain Pumir for fruitful discussions.
Finally, R. Zimmermann warmly thanks the TMB-2011 committee for assigning the ``Young Scientist Award'' of the Turbulence Mixing \& Beyond 2011 conference for the work he presented within this article.
\end{acknowledgments}

\bibliography{biblioJim}

\end{document}